\documentclass[aps,prl,twocolumn,groupedaddress]{revtex4}


\usepackage{graphicx,amsmath}

\begin{document}



\title{Detection of Single Electron Charging in an Individual 
InAs Quantum Dot by Noncontact Atomic Force Microscopy}


\author{Romain Stomp}
\author{Yoichi Miyahara}
\email[Corresponding author: ]{miyahara@physics.mcgill.ca}
\author{Sacha Schaer}
\author{Qingfeng Sun}
\author{Hong Guo}
\author{Peter Grutter}

\affiliation{Department of Physics, McGill University, 3600 rue University, Montreal, H3A2T8, Canada}

\author{Sergei Studenikin, Philip Poole, Andy Sachrajda}
\affiliation{Institute for Microstructural Science, National Research Council of Canada, Ottawa, K1AOR6, Canada }

\date{\today}

\begin{abstract}
Single electron charging in an individual InAs quantum dot 
was observed by electrostatic force measurements 
with an atomic force microscope (AFM). 
The resonant frequency shift and the dissipated energy of an oscillating AFM cantilever 
were measured as a function of the tip-back electrode voltage 
and the resulting spectra show distinct jumps when the tip was positioned above the dot. 
The observed jumps in the frequency shift, with corresponding peaks in dissipation,
are attributed to a single electron tunneling between the dot and the back electrode 
governed by Coulomb blockade effect,
and are consistent with a model based on the free energy of the system.
The observed phenomenon may be regarded as the ``force version'' 
of the Coulomb blockade effect.
\end{abstract}


\maketitle


Self-assembled semiconductor quantum dots (SAQDs) grown by lattice mismatched 
heteroepitaxy have attracted much attention as a promising system 
for many applications such as lasers, information storage devices and quantum computation.
There have been a considerable number of studies on the single electron charging effects 
on SAQDs located in field-effect structures 
because they enable the control of the charging state in the QDs by external electric fields.
These states can be probed by capacitance spectroscopy \cite{Drexler}
which provides information on the energy level structure
as well as the charging energy of the QDs \cite{Miller}.
 However, capacitance spectroscopy probes an ensemble of dots and
cannot be applied to an individual QD.
Access to individual QDs is considered to be a key technique not only 
for the further understanding of the physics of QDs, 
but also for some practical applications 
such as information storage and qubit read-out in quantum computation.

Scanning tunneling spectroscopy (STS) has been employed to investigate a single QD
\cite{Banin, Maltezopoulos}. 
 However, application of STS is limited to uncapped QDs on conducting substrates
since a tunneling current greater than 1 pA is usually required.
Electrostatic force measurement by atomic force microscopy (AFM) is known 
to have single electron sensitivity  \cite{Schoenenberger, Klein:2001,Klein:2002}. 
In these experiments, the observation of single electrons is based on the observation of 
quantized jumps in the force signal. 
Recently spectacular results on a QD incorporated in a carbon nanotube (CNT) were reported 
\cite{Woodside}. 
To determine single charging effects in these experiments, 
corroborating transport measurement through the QD via the CNT leads were necessary. 
This is unfortunately limiting for many interesting systems such as SAQDs or 
suspected charge traps leading to $1/f$ noise in mesoscopic devices, 
as contact leads cannot easily be attached. 
In this Letter, we report the observation of single electron charging events of 
a single SAQD by electrostatic force measurement
and present a simple theoretical model which explains
the main features of the experimental results.
As a consequence, optimal sample geometries can be designed 
and expected signal levels predicted 
for the experimental detection of single charging events.  
In addition we observe strong variations in the AFM force sensor damping,
which demonstrate the potential of this technique 
to investigate the fascinating interactions 
between micromechanical oscillators and single electron systems 
\cite{Schoelkopf, Knobel, Clerk}.  
\begin{figure}
 \includegraphics[width=7cm,keepaspectratio]{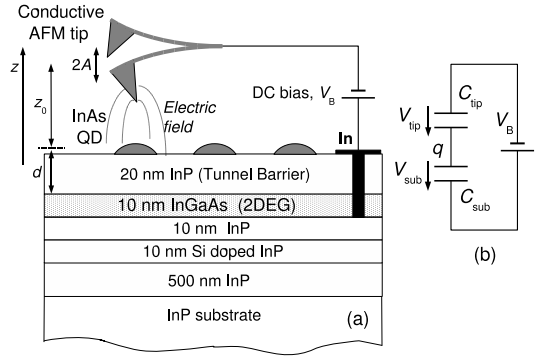}
 \caption{\label{setup} (a) Schematic diagram of the sample structure 
and the experimental setup, 
(b) equivalent electrical circuit, $q$ is the charge in the QD, 
$C_\mathrm{tip}$, $C_\mathrm{sub}$ are the tip-QD and
QD-backelectrode capacitances, respectively.}
\end{figure}

 The samples were prepared on a semi-insulating InP wafer by chemical beam epitaxy 
\cite{Lefebvre}. 
The schematic of the sample structure is depicted in Fig.~\ref{setup}.
The SAQDs spontaneously form due to lattice mismatched heteroepitaxy. 
A two dimensional electron gas (2DEG) formed in the InGaAs quantum well  
was used as a back electrode located 20 nm underneath the InAs SAQDs layer. 
 The electrical contact to the 2DEG was made by indium diffusion 
and low resistance Ohmic characteristics were confirmed between two such contacts at 4.2 K. 
The sample used in this experiment has a single layer of uncapped InAs SAQDs 
with a small dot density (5 $\mu {\rm m}^{-2}$ ). 
The typical dot is $50\pm10$ nm in diameter and 12 nm in height.
 In order to probe individual QD, we used an AFM in the frequency modulation mode
\cite{Albrecht}. 
\begin{figure}
 \includegraphics[width=8cm,keepaspectratio]{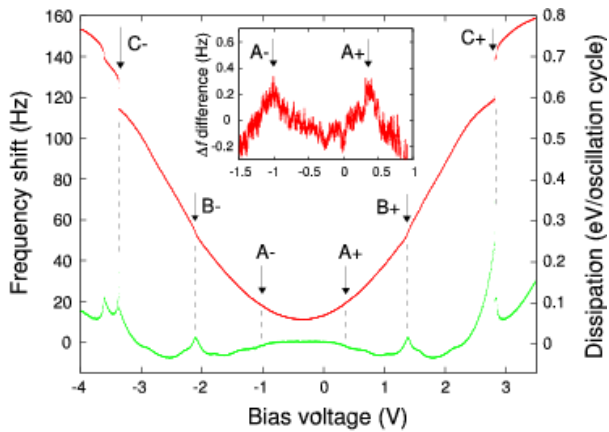}
 \caption{\label{df-v} Resonant frequency shift, $\Delta f$  and dissipated energy 
of the AFM cantilever as a function of the tip-sample bias voltage.
The arrows points the sudden increases in the $\Delta f$ caused by a single 
electron charging in a QD 
and they appear as a peak in the dissipation at the corresponding
bias voltage. 
The insets show the magnified spectrum around the structure A- and A+.
Here, a fitted parabola to the spectrum around the minimum is subtracted.}
\end{figure}
 In this technique, the AFM cantilever is self-oscillated at its mechanical resonance 
frequency $f_0$ by a positive feedback circuit with a phase shifter
and the resonance frequency shift $\Delta f$ caused by the tip-sample interaction is
measured by a phase locked loop \cite{Nanosurf}. 
 The oscillation amplitude, $A$, of the tip is held constant at 5 nm 
with an automatic gain controller (AGC). 
This enables the measurement of dissipation in the cantilever oscillation 
simultaneously to $\Delta f$.
 Our home-made cryogenic AFM \cite{Roseman:2000} uses a fiber-interferometric deflection 
sensor and has been previously used for successful imaging of vortices on Nb
\cite{Roseman:2001}.
The experiments were performed at 4.2 K in high vacuum of $1 \times 10^{-4}~{\rm mbar}$.
The cantilever used in our experiment had a resonance frequency, $f_0$, of 150 kHz 
with a spring constant, $k$, of 15 N/m.
 The tip was coated with a 10 nm Ti/ 20 nm Pt to ensure good electrical 
conductivity at 4.2 K.
After identifying a single QD by AFM by noncontact imaging, 
we performed series of electrostatic force spectroscopy (EFS) over the QD
as a function of tip-sample separation. 
 This spectroscopy records the change in the resonance frequency of the cantilever, 
$\Delta f$, caused by the tip-sample interaction as a function of the bias voltage 
between the tip and the back electrode while the distance regulation is turned off. 
The average tip-QD distance is typically more than 10 nm so that the tunneling between
the tip and the QD is negligible and the electrostatic force is the dominant interaction. 

Figure~\ref{df-v} shows a typical EFS spectrum and the dissipation signal.
The overall shape of the spectrum is characterized by a parabola which reflects
the capacitive force between the tip and the back electrode.
Since this force is attractive, the resonant frequency shift is negative.
For clarity, the negative frequency shift is plotted in all the following figures.
The minimum frequency shift at non-zero bias accounts for the contact potential
difference between the tip and the sample.
In addition to the parabolic background, some jumps are found in the frequency shift 
at various bias voltages.
We attribute them to the discrete change in the electrostatic force due to 
sequential charging of a single QD by a single electron
tunneling between the QD and the back electrode (Coulomb blockade).
These Coulomb blockade (CB) jumps are also observed in the dissipation signal
as peaks at the same bias voltages.
The increase in dissipation is obviously related to the dissipated energy
in the electron tunneling process.
This correspondence is helpful to identify the CB jumps at a lower bias voltage
whose frequency shift counterpart tends to be identified with difficulty.
No structure like those mentioned above were observed on the sample 
without the QD.

 We consider a simple model based on the free energy of the system 
as depicted in Fig.~\ref{setup}(b) to calculate the force acting on the AFM tip.
The free energy consists of the electrostatic charging energy and
the work done by the voltage source 
and can be expressed as \cite{Waser}:
\begin{equation}
W =  \frac{q^{2}}{2C_{\mathrm{\Sigma}}}-\frac{{C_{\mathrm{tip}}}}{C_{\mathrm{\Sigma}}}qV_{\mathrm{B}}
-\frac{1}{2}\frac{C_{\mathrm{sub}}C_{\mathrm{tip}}}{C_{\mathrm{\Sigma}}}V_{\mathrm{B}}^{2} . 
\label{free_energy}
\end{equation}
Here $q$ is the charge residing in the QD.
$C_\mathrm{sub}$ and $C_\mathrm{tip}$ are the QD-substrate, 
the tip-QD capacitance and
$C_{\mathrm{\Sigma}}=C_{\mathrm{tip}}+C_{\mathrm{sub}}$, respectively.
The force acting on the tip $F$ can be obtained by $F=-\partial W/\partial z$
where $z$ is the tip-QD distance.
Then we get
\begin{subequations}
\begin{eqnarray}
F&=&\frac{1}{C_{\mathrm{\Sigma}}^{2}}\frac{\partial C_{\mathrm{tip}}}{\partial z}\left( \frac{q^{2}}{2}
-C_{\mathrm{sub}}qV_{\mathrm{B}}+\frac{1}{2}C_{\mathrm{sub}}^{2}V_{\mathrm{B}}^{2}\right)
  \label{eq:force}\\
&=& \frac{1}{2}\frac{\partial C_{\mathrm{series}}}{\partial z}\left(V_\mathrm{B}-\frac{q}{C_\mathrm{sub}}\right)^2,\label{eq:force2}
\end{eqnarray}
\end{subequations}
where $C_{\mathrm{series}}=C_{\mathrm{tip}}C_{\mathrm{sub}}/(C_{\mathrm{tip}}+C_{\mathrm{sub}})$.
The first term accounts for the interaction 
between the charge in the QD and its image charge in the tip
but it is negligibly small under our experimental conditions.
The third term shows the parabolic background 
and accounts for the interaction between the polarized charges in the tip 
and the back electrode.
The interaction between the charge in the QD 
and the polarized charge in the tip is actually included in the second term
and is responsible for the detection of the charge in the QD.
It should be noticed from Eq.~(\ref{eq:force2}) that the expression reduces to 
a simple parabola when $q$ is independent of $V_\mathrm{B}$. 

\begin{figure}[b]
 \includegraphics[width=7.8cm,keepaspectratio]{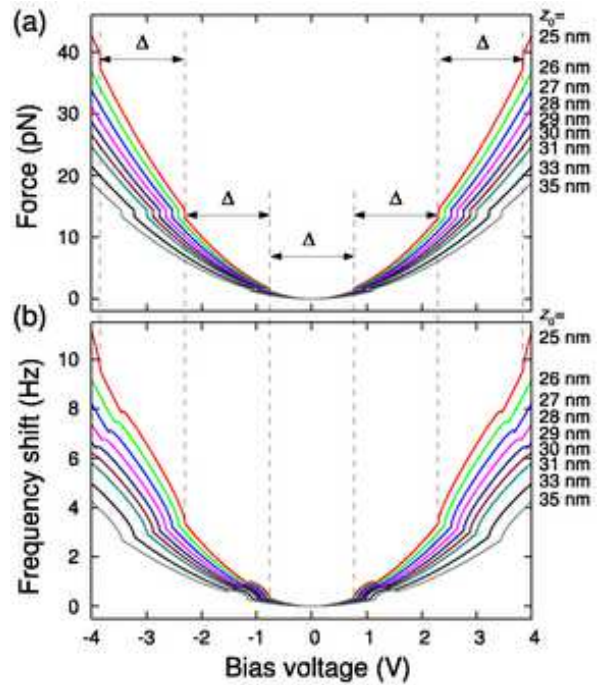}
 \caption{\label{calculated} Calculated (a) electrostatic force 
and (b) corresponding frequency shift as a function of $z_0$ 
using Eq.~(\ref{eq:force}), (\ref{eq:int}) and (\ref{eq:deltaf}).
A parallel plate capacitor model with an area of 227 $\mathrm{nm}^2$
(17 nm diameter disk) is assumed.}
\end{figure}

In this system, unlike the double tunneling junction which has been investigated by STS, 
only an electron tunneling between the back electrode and the QD is possible
because of the large tip-QD distance.
For this tunneling to be possible, the final state must be energetically favorable.
This requires $W(n+1)<W(n)$ for an electron to tunnel onto the QD with $n$ electrons,
and $W(n-1)<W(n)$ for an electron to tunnel off the QD with $n$ electrons. 
This determines the bias range (Coulomb blockade) in which the electron tunneling is forbidden:
\begin{equation}
  \label{eq:bias}
   \frac{e}{C_\mathrm{tip}}\left(n- \frac{1}{2}\right)< V_\mathrm{B} <\frac{e}{C_\mathrm{tip}}\left(n+ \frac{1}{2}\right).
\end{equation}
This translates into the condition, $-E_c/e < V_\mathrm{sub} < E_c/e=e/2C_\Sigma$
which relates the charging energy of the QD, $E_c$, to the applied voltage to the QD
through the relation, 
$V_\mathrm{sub}=(C_\mathrm{tip}V_\mathrm{B}-ne)/C_{\Sigma}$.
Eq.~\ref{eq:bias} leads to  
\begin{equation}
  \label{eq:int}
   q=-ne=-e~\mathrm{Int}\left(\frac{C_{\mathrm{tip}}V_{\mathrm{B}}}{e}+\frac{1}{2}\right)
\end{equation}
where the function $\mathrm{Int}$ gives the nearest integer to the argument.
By combining Eq.~(\ref{eq:force}) and Eq.~(\ref{eq:int}), the force can be obtained as a function
of the bias voltage.
The calculated $F$-$V_\mathrm{B}$ curves are shown in Fig.~\ref{calculated}(a)
for various $z_0$.
Step-like structures are found on the parabolic background.
The distance between two neighbouring jumps is constant 
and given by $\Delta =e/C_\mathrm{tip}$.
The step height increases at higher bias voltages 
because it is proportional to the $V_\mathrm{B}$ as can be seen in 
the second term of Eq.~(\ref{eq:force}).
This means that the structure nearer the zero bias is harder to observe.
A closer look at Eq.~(\ref{eq:force}) shows
that decreasing $C_{\mathrm{sub}}$ 
(increasing the distance between the QD and the back electrode)
enhances the jumps and reduces the parabolic background.
Note that increasing the QD-back electrode separation decreases the tunneling rate.

The resonant frequency shift of the cantilever measured in EFS 
is related to the force through the relationship \cite{Giessibl}:
\begin{equation}
  \label{eq:deltaf}
 \Delta f(z_{0}) = \frac{f_{0}^2}{kA}\int _{0}^{1/f_{0}}
  F(z_{0}+A~\mathrm{cos}(2\pi f_0t))\mathrm{cos}(2\pi f_0t)dt.
\end{equation}
The frequency shift is a weighted average of the force
over one oscillation period.
The calculated $\Delta f$-$V_{\mathrm B}$ curves are shown in Fig.~\ref{calculated}(b).
Although the step in $F$-$V_{\mathrm B}$ curve translates into broader increase in $\Delta f$
because of the averaging, 
the onset of the increase still corresponds to the step in $F$-$V_{\mathrm B}$ curve 
at the closest distance in one oscillation period.
This allows us to determine $C_\mathrm{tip}$ from $\Delta f$-$V_{\mathrm B}$ curves
using Eq.~(\ref{eq:bias}).
When we look at Fig.~\ref{df-v} carefully, the spacings between two neighboring
jumps are not exactly the same. 
One reason is that the oscillation amplitude decreases around the jumps due to
feedback errors of the AGC. 
The decrease in the amplitude leads to an increase in the closest tip-QD distance 
which results in the shift of the jumps to higher bias voltage.
The significant decrease in amplitude was actually observed at jump B and C, respectively.
The sharper increase at jump C is also due to the smaller amplitude.
The shift of the jump due to this effect should be corrected
in order to investigate the detail of the spectra,
such as internal energy levels of the QD 
where the separations between neighboring jumps are of serious concern. 
Regardless, we focus on the tip-QD distance dependence of the jump B
to demonstrate that the observed feature is consistent with
the theory discussed above.

\begin{figure}
 \includegraphics[width=8cm,keepaspectratio]{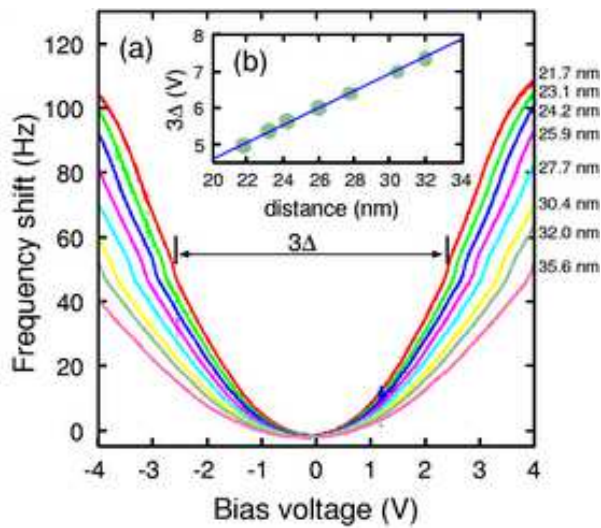}
 \caption{\label{zdepend} (a) Electrostatic Force Spectra 
as a function of the tip-QD distance. The number on the righthand side is
the absolute tip-QD distance obtained from a linear fitting.
(b)  Spacing of the jump B- and B+ versus
tip-QD distance.}
\end{figure}

Figure~\ref{zdepend}(a) shows a series of EFS spectra taken over a QD 
at various tip-QD distances.
As expected from the theory, the jump shifts to the lower bias voltage
as the distance becomes smaller because of the larger $C_\mathrm{tip}$.
Assuming that the spacing between B+ and B- is equal to $3\Delta$, 
$C_\mathrm{tip}$ is found to range from 0.064 to 0.094 aF.
This is one order of magnitude smaller than that in the STS experiment. 
In STS such a small value is not permissible 
because of associated low tunneling rate which is far less 
than that equivalent to a current of 1 pA.
In other words, the electrostatic force detection is sensitive 
to even a single electron charging event 
unlike STS measurements which statistically average a large number of such events.
Larger $\Delta f$ in the experiment than the calculated one is attributed to 
the electrostatic force between the tip and the substrate around the QD
which is not taken into account in the calculation. 
It also accounts for less sharp jumps in the experimental spectra.
 
As is shown in Fig.~\ref{zdepend}(b), 
the spacing between the jumps B+ and B-, 
$3\Delta$ is linearly dependent on the tip-QD distance.
This indicates $C_\mathrm{tip}\propto 1/z_0$ and 
it implies that the parallel plate capacitor model is valid in this
distance range.
Using a linear fitting of $\Delta$ versus distance plot, 
the absolute tip-QD distance and the effective area of the QD can be
determined.
The resulting distance ranges from 22 to 42 nm 
and the effective QD diameter is 17 nm.
The discrepancy between the effective and the measured diameter
is due to the parallel plate approximation of the lens shape QD  
as well as a depletion layer likely formed on the QD surface
by surface oxidation which affects the effective size of the QD.
These jumps and the corresponding peaks were also observed in $\Delta f$-$z$ curves 
and in the dissipation-$z$ curves at a fixed bias voltage (data not shown here). 
This can be understood by considering the change in $C_\mathrm{tip}$ along with
Eq.~(\ref{eq:int}) and it provides additional evidence 
for the observation of single electron effects.
The correlation of the peak in the dissipation with the jumps in the EFS spectra 
is also a good indication of the electron hopping on and off the QD 
with the oscillating tip.
Joule dissipation of moving charges has been reported previously \cite{Denk,Stowe},
but a quantitative calculation of the theoretically expected dissipation is
more than an order magnitude off.
We are presently investigating if the backaction 
of single electron charging events on the micromechanical oscillator 
can account for the observed dissipation.

In conclusion, we detected a single electron charging of an individual InAs QD 
by electrostatic force measurement. 
The observed features could be explained by a simple theory based on 
consideration of the free energy of the tip-QD-back electrode system. 
This theoretical understanding allows the optimization of sample geometries 
(in particular the back electrode to QD spacing).
This will enable experimental investigation 
of single charging events in diverse systems 
such as SAQDs and charge traps in mesoscopic systems. 
In contrast to STS, this technique can be used to investigating a QD 
only weakly coupled to an external electrode. 
Finally, we have observed strong contrast in dissipation, 
which cannot be explained by classical Joule dissipation. 
We currently only speculate that this is due to back action effects 
of single electron charging events on the micromechanical AFM oscillator.   

This research was supported by the Natural Science and Engineering Research
Council of Canada, the Canadian Institute of Advanced Research, and NanoQu\'ebec.

\end{document}